\documentclass{aastex63}

\usepackage{graphics}
\usepackage{float}
\usepackage{multirow}
\usepackage{makecell}

\received{June 1, 2019}
\revised{January 10, 2019}
\accepted{\today}
\submitjournal{Publications of the Astronomical Society of the Pacific}

\begin{document}

\title{First sodium laser guide star asterism launching platform in China on 1.8m telescope at Gaomeigu Observatory}

\correspondingauthor{You-Ming Guo}
\email{guoyouming@ioe.ac.cn}
\correspondingauthor{Min Li}
\email{limin\_alanna@163.com}

\author{Rui-Tao Wang}
\affiliation{Institute of Optics and Electronics, Chinese Academy of Sciences, Chengdu 610209, China}
\affiliation{University of Chinese Academy of Sciences, Beijing 100049, China}
\affiliation{Key Laboratory on Adaptive Optics, Chinese Academy of Sciences, Chengdu 610209, China}

\author{Hong-Yang Li}
\affiliation{CAS Key Laboratory of Optical Astronomy, National Astronomical Observatories, Chinese Academy  of Sciences, Beijing 100012, China}
\affiliation{University of Chinese Academy of Sciences, Beijing 100049, China}

\author{Lu Feng}
\affiliation{CAS Key Laboratory of Optical Astronomy, National Astronomical Observatories, Chinese Academy  of Sciences, Beijing 100012, China}

\author{Min Li}
\affiliation{Institute of Optics and Electronics, Chinese Academy of Sciences, Chengdu 610209, China}
\affiliation{Key Laboratory on Adaptive Optics, Chinese Academy of Sciences, Chengdu 610209, China}

\author{Qi Bian}
\affiliation{ Technical Institute of Physics and Chemistry, Chinese Academy of Sciences, Beijing 100190, China}

\author{Jun-Wei Zuo}
\affiliation{ Technical Institute of Physics and Chemistry, Chinese Academy of Sciences, Beijing 100190, China}

\author{Kai Jin}
\affiliation{Institute of Optics and Electronics, Chinese Academy of Sciences, Chengdu 610209, China}
\affiliation{Key Laboratory on Adaptive Optics, Chinese Academy of Sciences, Chengdu 610209, China}

\author{Chen Wang}
\affiliation{ Technical Institute of Physics and Chemistry, Chinese Academy of Sciences, Beijing 100190, China}

\author{Yue Liang}
\affiliation{ Technical Institute of Physics and Chemistry, Chinese Academy of Sciences, Beijing 100190, China}

\author{Ming Wang}
\affiliation{ Technical Institute of Physics and Chemistry, Chinese Academy of Sciences, Beijing 100190, China}

\author{Jun-Feng Dou}
\affiliation{Institute of Optics and Electronics, Chinese Academy of Sciences, Chengdu 610209, China}
\affiliation{Key Laboratory on Adaptive Optics, Chinese Academy of Sciences, Chengdu 610209, China}

\author{Ding-Wen Zhang}
\affiliation{Institute of Optics and Electronics, Chinese Academy of Sciences, Chengdu 610209, China}
\affiliation{Key Laboratory on Adaptive Optics, Chinese Academy of Sciences, Chengdu 610209, China}

\author{Kai Wei}
\affiliation{Institute of Optics and Electronics, Chinese Academy of Sciences, Chengdu 610209, China}
\affiliation{Key Laboratory on Adaptive Optics, Chinese Academy of Sciences, Chengdu 610209, China}

\author{You-MIng Guo}
\affiliation{Institute of Optics and Electronics, Chinese Academy of Sciences, Chengdu 610209, China}
\affiliation{Key Laboratory on Adaptive Optics, Chinese Academy of Sciences, Chengdu 610209, China}

\author{Yong Bo}
\affiliation{ Technical Institute of Physics and Chemistry, Chinese Academy of Sciences, Beijing 100190, China}
and\author{Sui-Jian Xue}
\affiliation{CAS Key Laboratory of Optical Astronomy, National Astronomical Observatories, Chinese Academy  of Sciences, Beijing 100012, China}



\begin{abstract}

The application of sodium laser guide star is the key difference between modern adaptive optics system and traditional adaptive optics system. Especially in system like multi-conjugate adaptive optics, sodium laser guide star asterism which is formed by several laser guide stars in certain pattern is required to probe more atmospheric turbulence in different directions. To achieve this, a sodium laser guide star asterism launching platform is required. In this paper, we will introduce the sodium laser guide star asterism launching platform built and tested on the 1.8m telescope of the Gaomeigu Observatory. The platform has two functions: one is to compare the performance of sodium laser guide stars generated by different lasers at the same place; the other is to generate sodium laser guide star asterism with adjustable shape. The field test results at the beginning of 2021 verify the important role of the platform, which is also the first time to realize sodium laser guide star asterism in China.

\end{abstract}



\section{Introduction} \label{sec:introduction}
In order to eliminate the influence of atmospheric turbulence, adaptive optics systems utilizing sodium laser guide stars have gradually become one of the necessary subsystems of large ground-based optical-infrared telescopes, such as Keck Telescope (\citealt{Chin+etal+2016}), Gemini Telescopes (\citealt{Fesquet+etal+2013}), the European Southern Observatory’s Very Large Telescope (VLT, \citealt{Calia+etal+2014}) \textit{et cetera}. In order to enable tomographic sensing of atmospheric aberration and expand the correction field of view of single sodium laser guide star adaptive optics system, it is necessary to use multiple laser generated artificial guide stars to form different constellation structures, which is more important for larger aperture telescopes. 

At present, there are mainly two effective methods for generating sodium laser guide star asterism that have already been practiced on telescopes. One of such methods is to utilize multiple laser heads. Small telescopes with diameter around 30cm are installed before each laser head forming several separated laser guide star launching units. By individually adjusting the laser head's power output and the launching telescope's pointing direction of each unit, it is possible to form laser guide star asterism with different power distribution and shape. VLT's Laser Guide Star Facility (LGSF) is an example of such configuration utilizing four sodium laser heads and four launching telescopes. Although this configuration has the advantage of freedom for controlling each individual laser guide star's performance, the cost of such system is relatively high and the adjustment of each subsystem can be complex. The other method is to use a single high-power laser head for generating sodium laser, and splitting the output laser into multiple beams and using a optics setup called asterism generator before one single launching telescope to generate laser guide star asterism. One successful example of such configuration is Gemini south Telescope's GeMS/GSAOI system (\citealt{Dalessandro+etal+2016}). By reducing the number of launching telescopes and laser heads, the cost can be lower and it is much easier to installed the single launching telescope behind the secondary mirror of the telescope for central launch, however, because all beams have to be generated with a single laser head, the power of the laser have to be particularly large. The design, manufacturing and even adjusting such a laser can be difficult, especially for long term maintenance. 

In China, the largest existing general optical/infrared astronomical telescopes in China are still 2 meter class telescopes by this moment. Before the advent of larger optical infrared astronomical telescopes, the community would like to test out adaptive optics system with multiple laser guide stars on such telescopes. In this situation, the LGSF's configuration using multiple launching units installed on the side of primary mirror cage would be too heavy for the telescope even using our compact 30W sodium laser head. 

To solve this problem, we have developed a sodium laser guide star asterism launching platform. The platform can not only generate configurable laser guide star asterism, but could also be used to simultaneously compare on sky performances of lasers installed on the platform. In this paper, we will introduce the laser guide star asterism launching and testing platform realized in the Lijiang 1.8m telescope in detail in Section \ref{section:Setup of the launching and testing platform}, and briefly introduce the first laser guide star asterism experiment in China and laser guide star performance calibration experiment after the successful construction of this platform in Section \ref{section: Laser guide star asterism and guide star performance calibration experiment}.

\section{Setup of the launching and testing platform}\label{section:Setup of the launching and testing platform}
\subsection{Receiving telescope system and detector for field test}
The receiving telescope for the field test is a 1.8m telescope developed by the Institute of Optics and Electronics,
\begin{figure}[H]
\centering
\includegraphics[width=5cm, height=10cm]{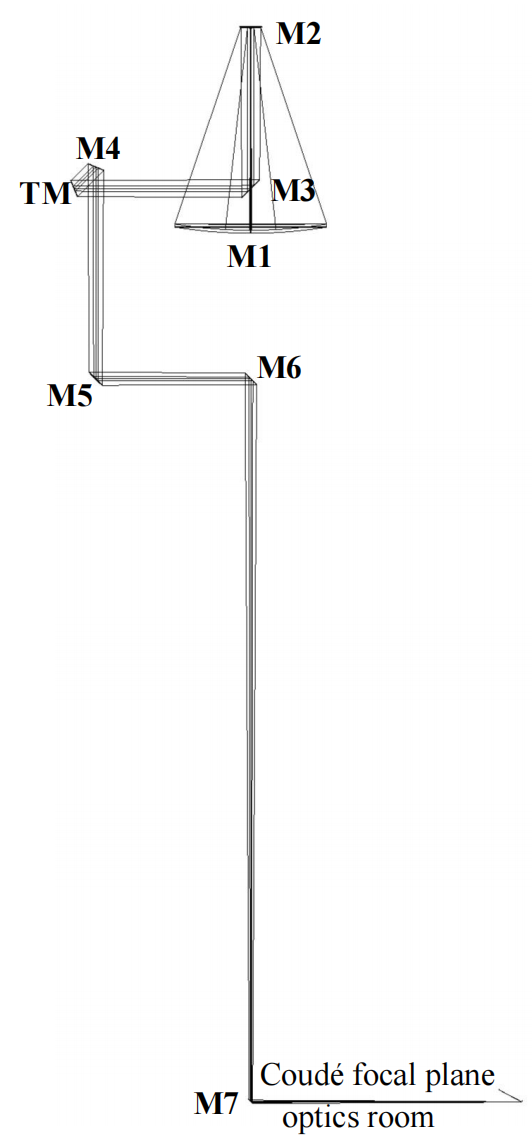}
\caption{The coude optical path layout of the 1.8m telescope}
\label{figure: Optical path layout of the 1.8m telescope}
\end{figure}
Chinese Academy of Sciences(IOE). The telescope is used for continuous imaging the launched sodium laser guide star asterism and monioring each individiual star's brightness and shape. The telescope is a typical Cassegrain optical system and is the best astronomical telescope applying adaptive optics in China (\citealt{Wei+etal+2010}). The main Coud\'e optical path layout of the 1.8m telescope is shown in Figure \ref{figure: Optical path layout of the 1.8m telescope}. The primary mirror M1 is a paraboloid with a focal ratio of F/1.5, and its clear diameter is 1760 mm. The F number of Coud\'e optical system is 89 and the design field of view is ± 1.5'. M2 represents the hyperboloid secondary mirror of the telescope, TM represents the fine tracking tip-tilt mirror, and M3-M7 is the reflect mirror of the Coud\'e optical path of the telescope. After the incident light passes through M1 and M2, the mirror M3 reflects the light beam to the fine tracking tip-tilt mirror TM to improve the tracking accuracy of the system. Then the mirror M4-M6 guides the light into the Coud\'e room, and finally the M7 mirror guides the light into the adaptive optics system in the Coud\'e room.

\begin{figure}[H]
\centering
\includegraphics[width=0.8\linewidth]{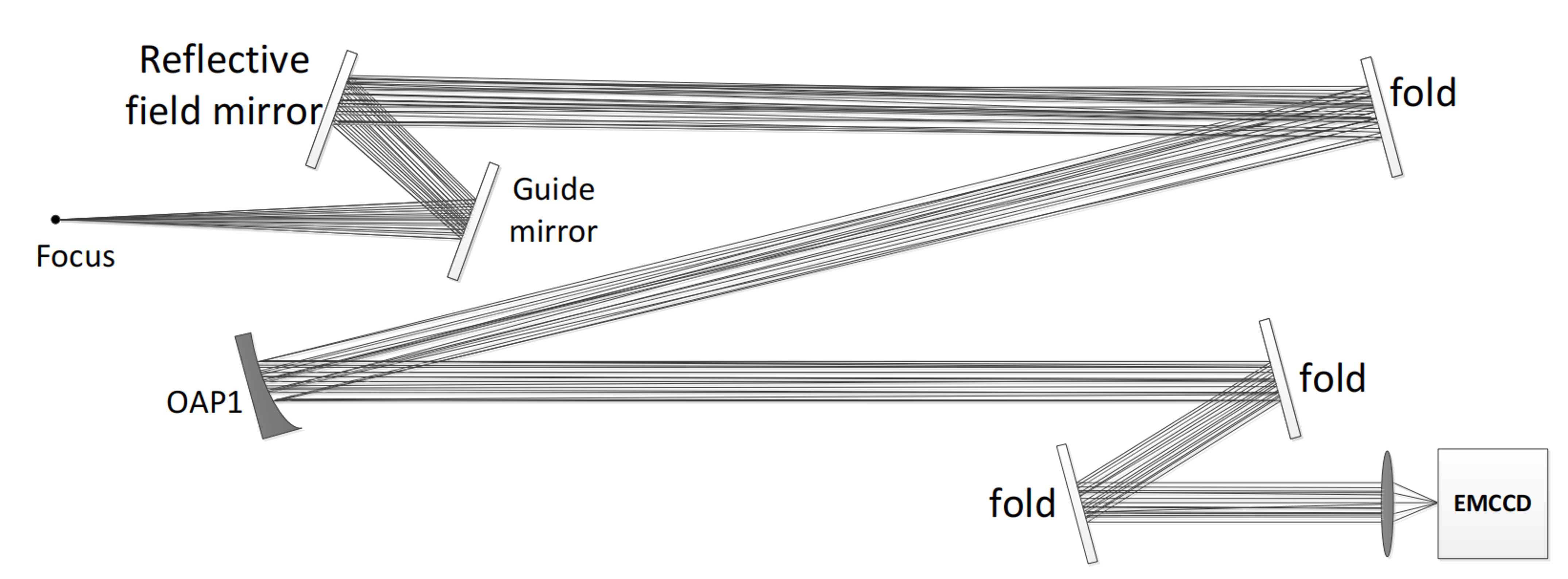}
\caption{The optical path in the Coude room}
\label{figure: the optical path in the Coude room}
\end{figure}

After passing through the main optical path (M1-M6) of the telescope, the return light of the sodium beacon enters the optical path of the Coud\'e room through the reflection of M7 mirror. Figure \ref{figure: the optical path in the Coude room} shows the optical path in the Coud\'e room. The return light will eventually enter the fine tracking detector after being reflected by multiple optical components. The fine tracking detector is an Andor iXon + ultra888 series EMCCD camera, and parameters of the detector are listed in Table \ref{table: the main parameters of the detector}.

\begin{table}[H]
\centering
\begin{tabular}{|c|c|}
\hline
Parameters                            & Value                                  \\ \hline 
pixels (H $\times$ V)                 & 1024 $\times$ 1024                     \\ \hline
Pixel size (W $\times$ H; $\mu m$)    & 13 $\times$ 13                          \\ \hline
Image area (mm)                       & 13.3 $\times$ 13.3                      \\ \hline
Pixel Well Depth ($e^-$)              & 80,000                                 \\ \hline
Max Readout Rate (MHz)                & 30                                     \\ \hline
Frame rates (fps)                     & 26 (full frame) - 9690                 \\ \hline 
Read noise ($e^-$)                    & $<$1 with EM gain                        \\ \hline
QE Max                                & $>$90\%                                   \\ \hline
Digitization                          & 16-bit                                 \\ \hline
PC Interface                          & USB 3.0                                \\ \hline
Timestamp accuracy                    & 10 ns                                   \\ \hline
Field of view (arcsec)                & $512 \times 512$                        \\ \hline
Single pixel field of view (arcsec)   & 0.5                                      \\ \hline
\end{tabular}
\caption{The main parameters of the detector}
\label{table: the main parameters of the detector}
\end{table}

\subsection{Laser guide star facility}
\begin{figure}[H]
\centering
\includegraphics[width=0.9\linewidth]{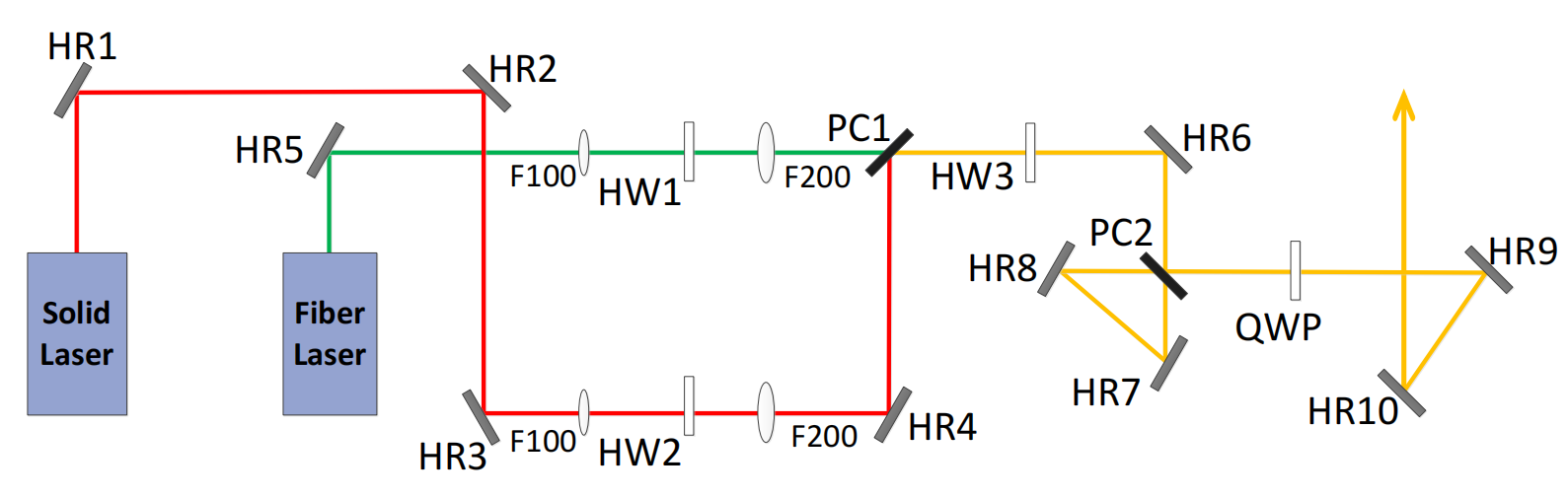}
\caption{The optical path layout of the laser guide star facility(HW is half wave plate; HR is high reflection mirror; PC is a polarizer; QWP is a quarter wave plate)}
\label{figure: the optical path layout of the laser guide star facility}
\end{figure}

The laser guide star facility(LGSF) is responsible for generating and launching the sodium laser guide star asterism. It includes sodium laser heads, beam transfer optics, a asterism generator and a laser launching telescope. The laser guide star facility is placed on a stationary platform on the side of the rotating azimuthal part of the telescope's mount. Figure \ref{figure: the optical path layout of the laser guide star facility} shows the optical path layout of this facility. On the left are two 30W sodium guide star lasers (solid-state and fiber sodium lasers). First, the two lasers output laser beams into their respective optical paths, and then adjust their polarization state to the horizontal direction (p polarization) and vertical direction (s polarization) respectively through the half wave plates HW1 and HW2. The polarization incoherent combination of two orthogonal polarized lasers is realized by the polarization synthesizer PC1 with the characteristics of high reflection of s-polarized light and high transmission of p-polarized light. The combined beams have a certain interval and the beam quality remains unchanged. Half wave plate HW1/HW2 and polarizer PC1 are combined as power attenuators to control the output power of two laser beams respectively. Place the half wave plate HW3 behind the polarizer PC1, the polarization direction of each laser beam will change after passing through HW3, and then split into two beams by the polarization beam splitter PC2; Among them, the vertical component is reflected by P2, and the horizontal component passes through P2, and then passes through P2 again after being reflected by the high mirror HR7 and HR8. The energy ratio between beams can be adjusted by rotating the angle between the HW3 axis and the x-axis.Finally, the four partially overlapping yellow laser beams are reflected and injected into the 300mm laser launch telescope(LLT) by the high reflection mirror HR10.The laser beam is expanded by LLT from 10mm to 240mm.The LLT has a field of view of  ± 2.5' and focuses the laser beam at 90km$\sim$210km.The total throughput of the LLT is around $90\%$(\citealt{Wei+etal+2016}).The piezoelectric ceramic actuator is used to precisely change the placement angle and position of the high reflection mirrors HR7, HR8, HR9 and HR10, optimize the included angle and separation distance between the optical axes of the four laser beams, and project them into the sky to form a laser guide star constellation. 

In order to further improve the photon return flux of the sodium guide star, a QWP is inserted in front of HR9, and the s-polarized light and p-polarized light are adjusted to circularly polarized state to obtain efficient pumping of sodium atoms.These setup could be used to form $2^N$ laser guide star by replicating all the optics from the last half-wave plate and PC in figure \ref{figure: the optical path layout of the laser guide star facility} into each beamlet's train if more powerful lasers are employed. At the moment of our test, our two lasers are both only 30W, therefore, we divide each laser into two beamlets to form two sodium guide stars, so into total we can form up to four sodium guide stars.

This facility can also be used to combine the two lasers to form a single sodium guide star with higher brightness, which is also a method to effectively improve the brightness of sodium guide star in case that the sodium column density in atmosphere is low or when the telescope is pointing at low altitude angle. In addition, the platform can also simultaneously test the performance of sodium guide star lasers generated by lasers of different formats. The results of such simultaneous testing will be free from temporal environmental differences like seeing, atmospheric transparency and sodium layer's dynamic and thus will be more easy to tell which laser will be better for generating sodium laser guide star.

\subsection{30W class sodium guide star lasers}
\begin{figure}[H]
\centering
\includegraphics[width=0.6\linewidth]{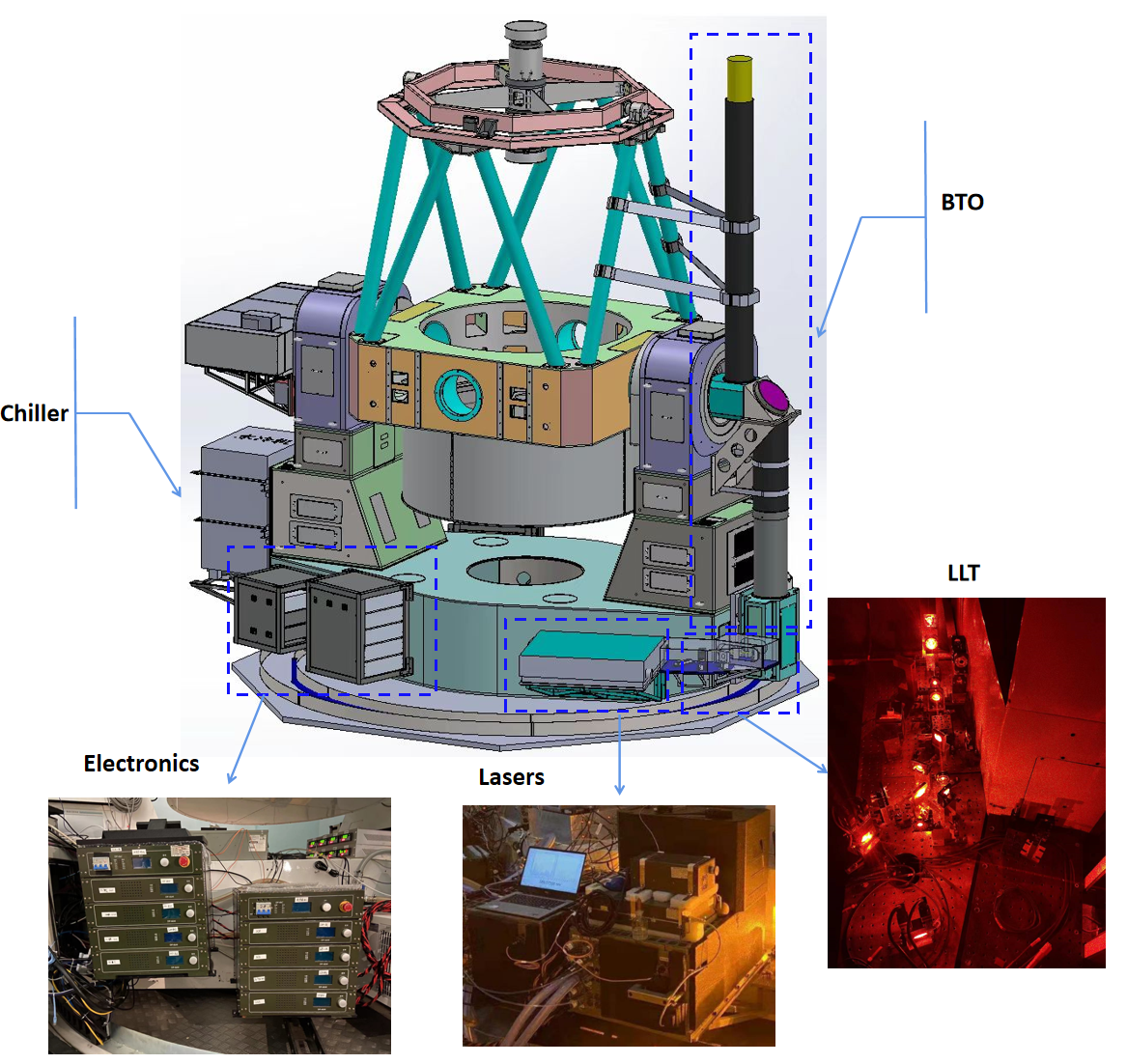}
\caption{The schematic diagram of the location of the LGS facility}
\label{figure: the schematic diagram of the location of the LGS facility}
\end{figure}

\begin{figure}[H]
\begin{minipage}[t]{0.5\linewidth}
\centering
\includegraphics[width=\linewidth]{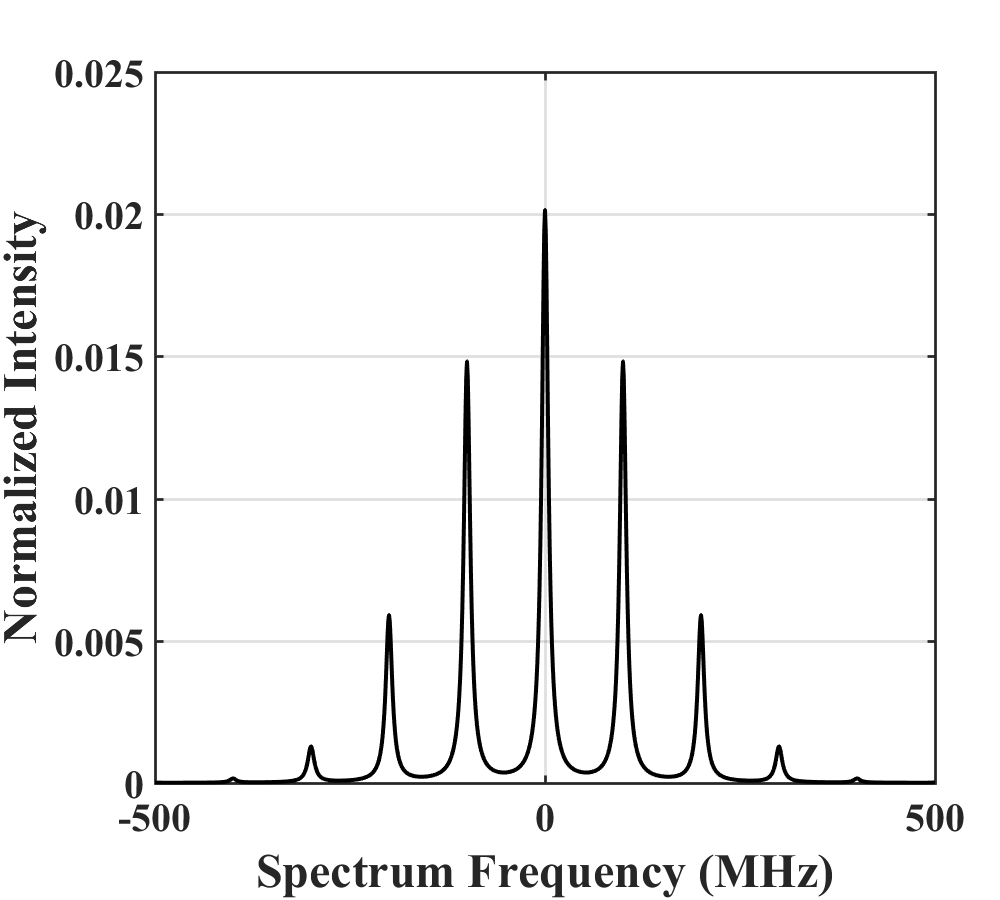}
\end{minipage}%
\begin{minipage}[t]{0.5\linewidth}
\centering
\includegraphics[width=\linewidth]{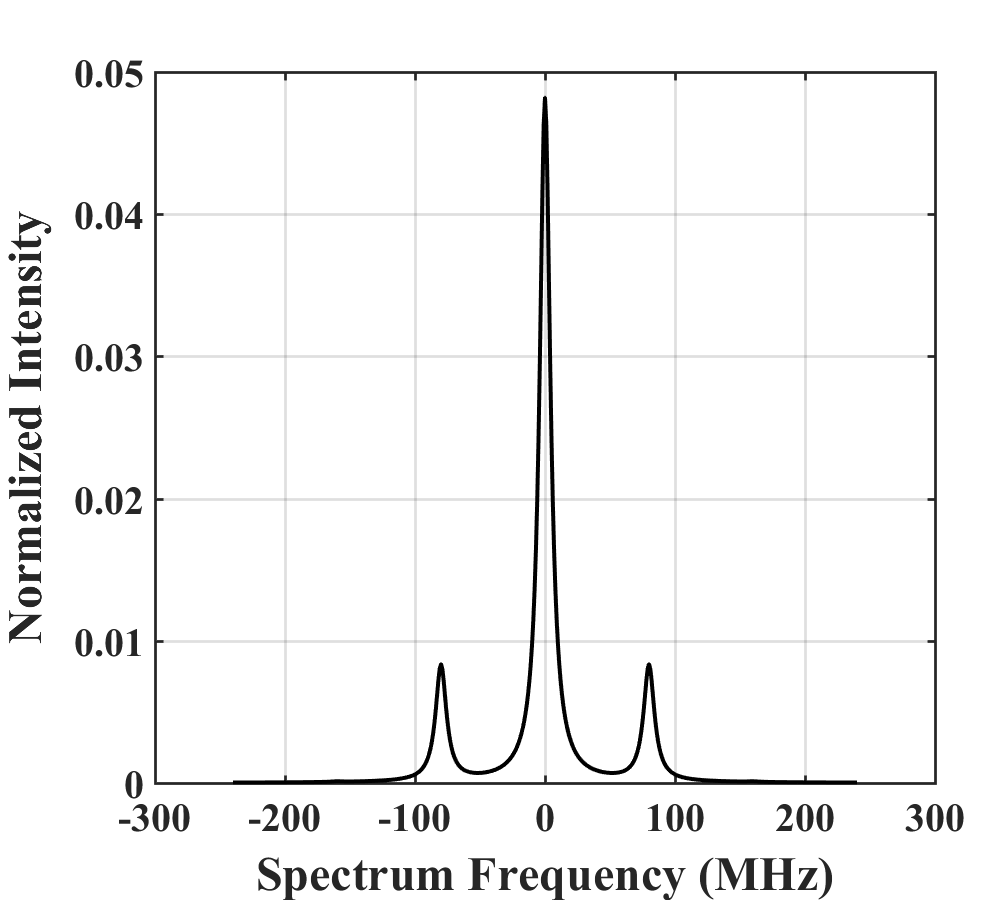}
\end{minipage}
\caption{Spectrum formats of the two TIPC 30W class QCW pulsed sodium lasers. (left) The width (FWHM) of spectrum is 300MHz and the longitudinal mode interval is 100 MHz for the all solid-state sodium laser.  (right) The width (FWHM) of spectrum is 100MHz and the longitudinal mode interval is 80 MHz for the fiber sodium laser.}
\label{figure: spectrum formats for the two sodium lasers}
\end{figure}

In Figure \ref{figure: the schematic diagram of the location of the LGS facility}, both lasers are 30W class quasi-continuous-wave(QCW) pulsed lasers developed by Technical Institute of Physics and Chemistry, Chinese Academy of Sciences (TIPC). The main difference of these lasers is their linewidth due to their different seed lasers. Both lasers could output 589 nm yellow beam by sum-frequency with 1064 nm 
and 1319 nm seed lasers. Compared with the all solid-state laser, the new laser has
two upgrades. One is that the 1064 nm solid-state seed laser was replaced by a fiber-seed laser, we will mention this laser as fiber sodium laser for convenience. The other is that because of this new seed laser, the linewidth of fiber sodium laser is narrower compared with the all solid-state sodium laser. The spectrum formats of the two lasers is shown in Figure \ref{figure: spectrum formats for the two sodium lasers}. The linewidth (FWHM) of spectrum is 300MHz and the longitudinal mode interval is 100 MHz for the all solid-state sodium laser (Figure \ref{figure: spectrum formats for the two sodium lasers} left). For the fiber sodium laser (Figure \ref{figure: spectrum formats for the two sodium lasers} right), the linewidth (FWHM) is 100MHz and the longitudinal mode interval is 80MHz. The beam quality(M2) of the two lasers is 1.4 and the more details of the parameters of the two sodium lasers are summarized in Table \ref{table: laser parameters list}.

\begin{table}[H]
\centering
\begin{tabular}{|l|l|l|}
\hline
\textbf{Parameter name}                                                                             & \textbf{All solid-state sodium laser}                                                               &
\textbf{Fiber sodium laser}\\ \hline
Central wavelength                 & $589.159$ nm            &            $589.159$ nm           \\ \hline
Linewidth                         & $\sim300$ MHz           &            $\sim100$ MHz          \\ \hline
Longitudinal mode interval        & $\sim100$ MHz           &            $\sim80$ MHz           \\ \hline
Longitudinal mode width (FWHM)    & $\sim10$ MHz            &            $\sim10$ MHz           \\ \hline
Pulse repetition rate             & $800$ Hz                &            $800$ Hz               \\ \hline
Pulse width                       & $\sim100$ $\mu$s        &            $\sim100$ $\mu$s       \\ \hline
Beam quality (M2)                 & 1.4                     &            1.4                    \\
\hline
Polarization                      & Circular                &            Circular               \\ \hline
Average power                     & $31$ W                  &            $30$ W                 \\ \hline
\end{tabular}
\caption{Laser parameters for the two 30W class QCW pulsed sodium lasers}
\label{table: laser parameters list}
\end{table}

\section{Laser guide star asterism and guide star performance calibration experiment}\label{section: Laser guide star asterism and guide star performance calibration experiment}
The platform was fabricated and installed onto 1.8m telescope in January 2021. It is unique in two aspects: 1. when launching two sodium laser on the sky within close range at the same place using the platform, the performance of the two lasers can be easily compared with differential photometry.Compared with the previous tests when only one laser can be launched every time
(\citealt{Wei+etal+2012,Jin+etal+2015}), influences from temporal variations of experimental parameters (such as sodium column density, seeing, transparency) and pointing direction of the telescope before and after replacement of another laser can be minimized as mentioned.; 2. it can not only combine the two laser beams to generate a brighter sodium guide star, but also generate laser guide star asterism with up to four sodium guide stars by using the two lasers. 

Using the newly built platform, we conducted field tests for generating laser guide star asterism and guide star performance calibration on the two afore-mentioned 30W QCW pulsed sodium lasers at Gaomeigu site in Lijiang, Yunnan Province from December 29, 2020 to January 15, 2021. There are two main reasons for choosing this time to conduct experiments on sodium guide stars. One is that the sodium column density is the highest at Gaomeigu site in this period of year (\citealt{Li+etal+2021}), and the generated sodium guide star will have higher photon return flux compared with other months. The other reason is that Gaomeigu site has a relatively high proportion of clear nights in a year.Throughout the test, a standard Johanson V band optical filter is always placed in front of the CCD, and differential photometry(\citealt{Feng+etal+2016}) is used.

\subsection{Test on laser guide star performance of two types of lasers}
\label{subsection: Test on laser guide star performance of two types of lasers}
Since the two lasers employed have different linewidth, to answer the question whether a narrower or broader linewidth would be better suited for sodium laser guide star generation would be perfect test application for this platform. 

\begin{figure}[H]
\centering
\includegraphics[width=\linewidth]{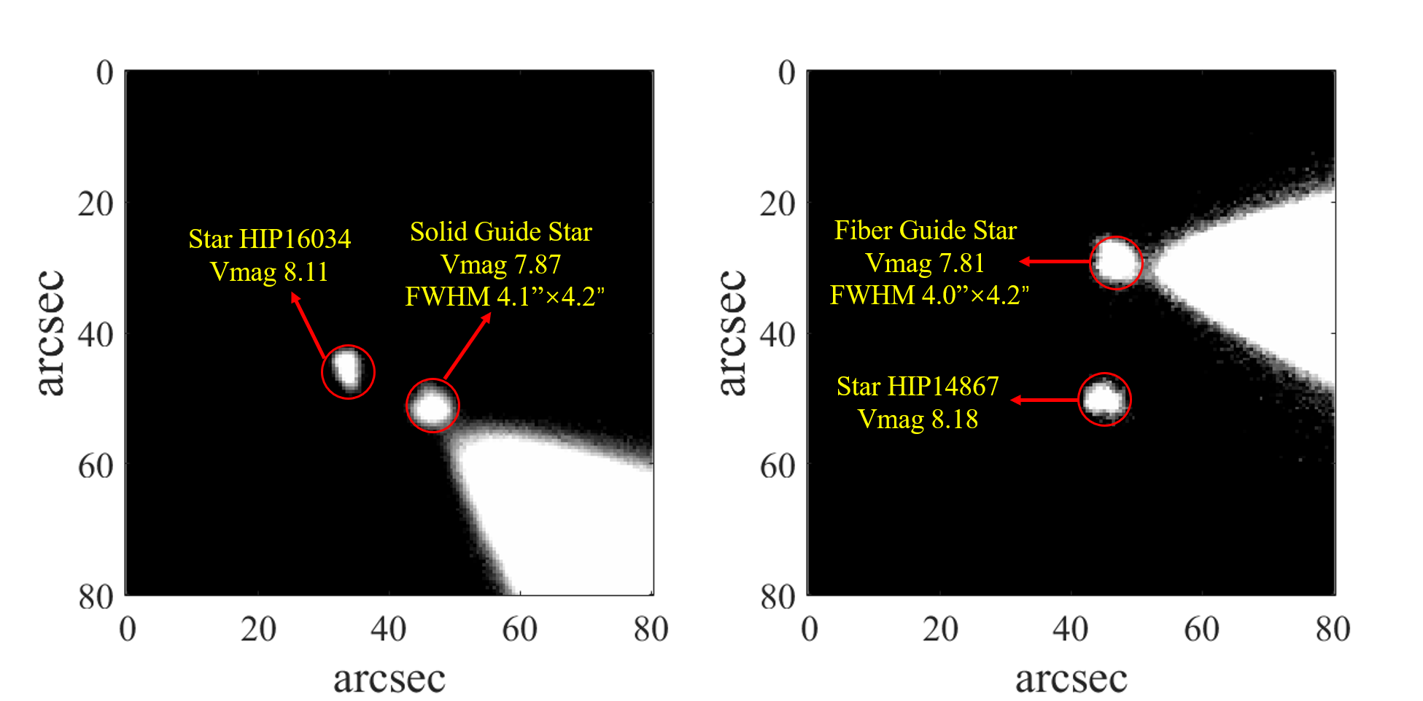}
\caption{The guide star of the all solid-state sodium laser and star HIP16034 with brightness of V8.11 magnitude(left). The guide star of the fiber sodium laser and star HIP14867 with brightness of V8.18 magnitude(right)}
\label{figure: guide star of the two lasers}
\end{figure}

Before the experiment, the optimal wavelength of the two lasers is needed to be determined. The wavelength was gradually adjusted from small to large and it was the optimum wavelength of the laser when the brightness of the guide star reaches the highest. Then the optimal polarization state is needed to be determined. The angle of the polarizer(QWP in Figure \ref{figure: the optical path layout of the laser guide star facility}) was gradually adjusted, when the brightness of the guide star was the highest, it was exactly the position of circular polarization. In the experiment, the power of both lasers was adjusted to a maximum of 30W. We conducted experiments on the performance of two lasers on the same day. Figure \ref{figure: guide star of the two lasers} left shows the guide star of the all solid-state sodium laser and star HIP16034 with brightness of V8.11 magnitude. The brightness of the solid guide star is 7.87 magnitude in V band through differential photometry and the on-sky spot size (FWHM) of the solid guide star is $4.1” \times 4.2”$. Figure \ref{figure: guide star of the two lasers} right shows the guide star of the fiber sodium laser and star HIP14867 with brightness of V8.18 magnitude. The brightness of the fiber guide star is 7.81 magnitude in V band through differential photometry and the on-sky spot size (FWHM) of the fiber guide star is $4.0” \times 4.2”$.

\subsection{Combined laser guide star and laser guide star asterism experiment}
\label{subsection: Combined laser guide star and laser guide star asterism experiment}
\begin{figure}[H]
\centering
\includegraphics[width=0.6\linewidth]{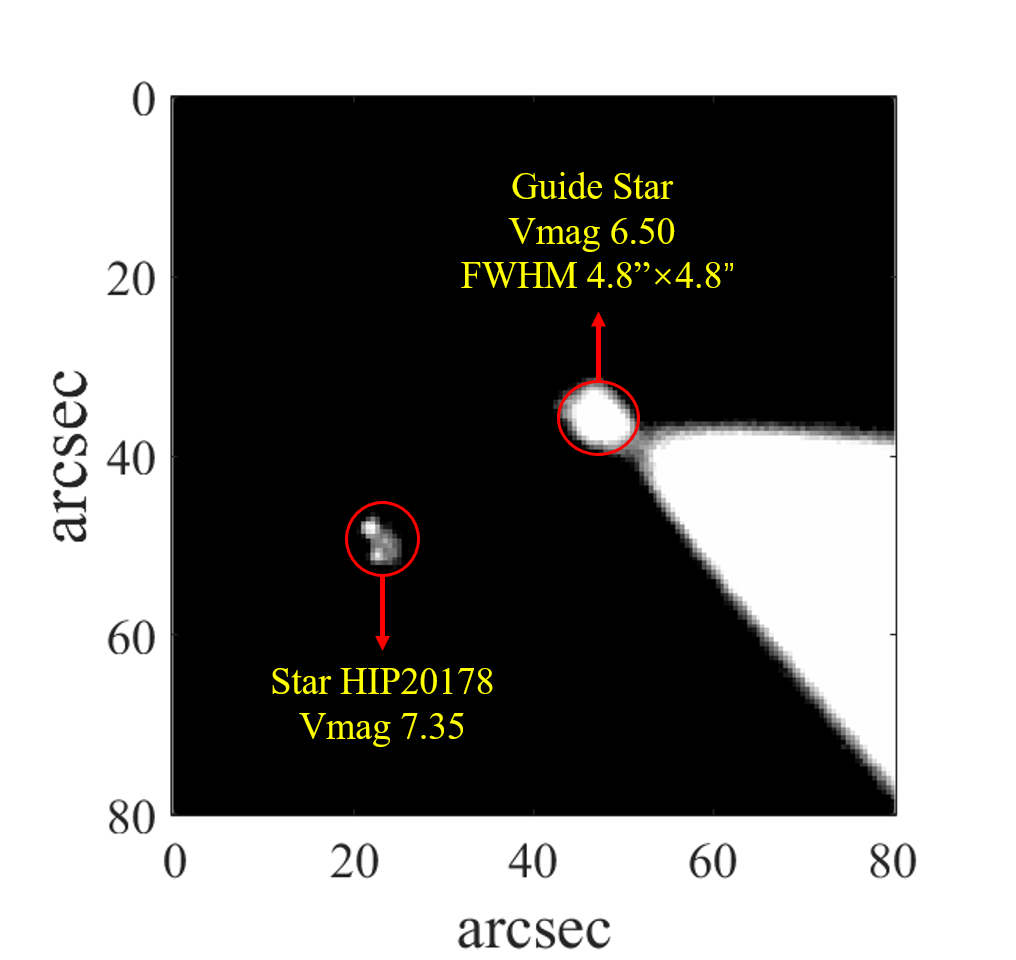}
\caption{The combined guide star and star HIP20178}
\label{figure: combined guide star}
\end{figure}

The beams of the all solid-state sodium laser and the fiber sodium laser can combined into one single beam by using the laser guide star facility shown in Figure \ref{figure: the optical path layout of the laser guide star facility}  and generate a bright guide star on sky. Figure \ref{figure: combined guide star} shows the sodium laser guide star formed by this beam combination method. A reference star HIP20178 with brightness of 7.35 magnitude in V-band is also in the frame. The field of view of this figure is $80arcsec \times 80arcsec$. During this experiment, the seeing of the Gaomeigu site was 2.5". The brightness of the combined guide star was 6.5 magnitude in V band and the size of the guide star(FWHM) was $4.8" \times 4.8"$.

\begin{figure}[H]
\centering
\includegraphics[width=0.6\linewidth]{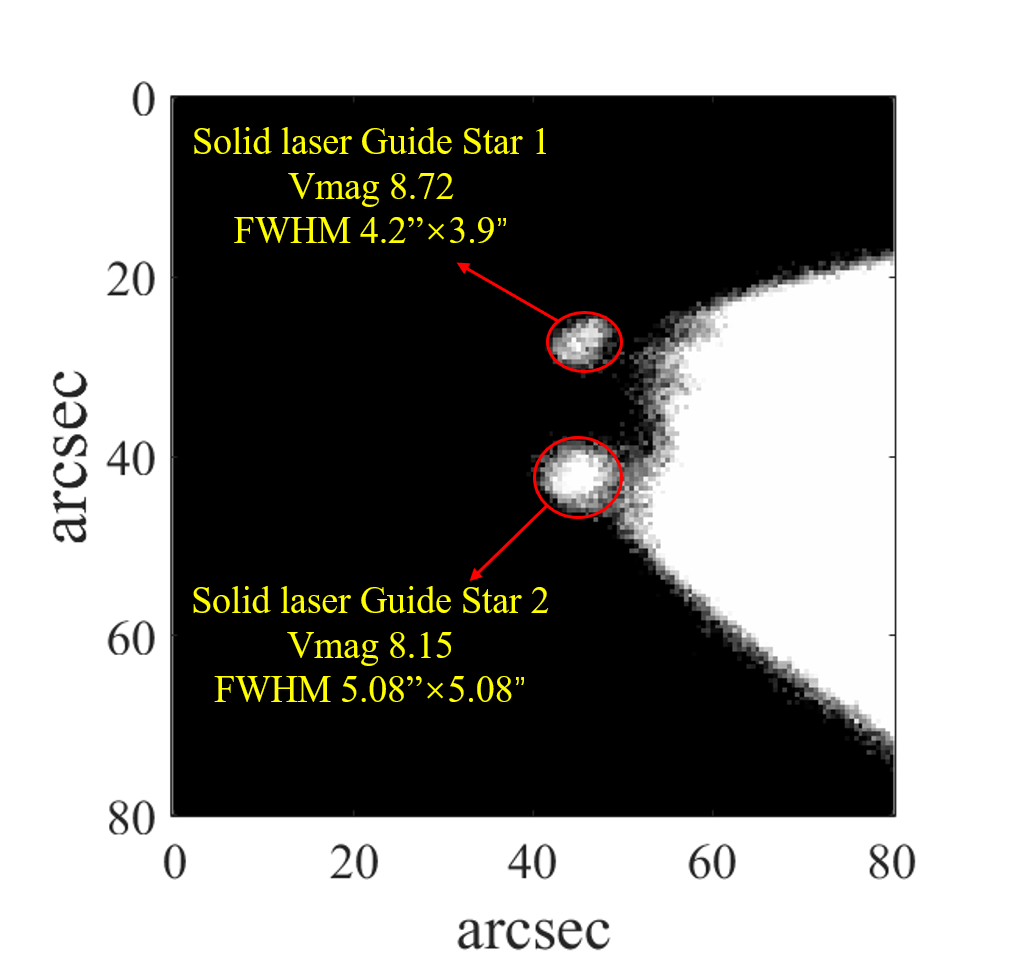}
\caption{The two guide stars of the all solid-state sodium laser}
\label{figure: the two guide stars of the all solid-state sodium laser}
\end{figure}

Using the laser guide star facility, a single sodium laser can generate two guide stars and each star’s power could be controlled individually and the shape of asterism could be adjusted. Taking the all solid-state sodium laser as an example, as shown in Figure \ref{figure: the two guide stars of the all solid-state sodium laser}, the light of the solid-state laser was divided into two beams and projected onto the sodium layer to generate a close guide stars pair. With a local seeing of 2.5", the brightness of the solid laser guide star 1 and 2 are V8.72 mag and V8.15 mag, respectively, and the size of the guide star (FWHM) 1 and 2 on sky are $4.2" \times 3.9"$ and $5.08" \times 5.08"$. The parameters are summarized in Table \ref{table: the parameters of double guide star experiment of all solid-state laser}.
\begin{table}
\scalebox{0.7}
\centering
\begin{tabular}{|l|l|l|}
\hline
\textbf{Parameter name}                                                                             & \textbf{Solid laser guide star 1}                                                               &
\textbf{Solid laser guide star 2}\\ \hline
Projected direction     &  Zenith               &  Zenith                \\ \hline
Seeing                  &  2.5"                 &  2.5"                  \\ \hline
Brightness              & 8.72 Vmag             & 8.15 Vmag              \\ \hline
The Spot size           & $4.2" \times 3.9"$    & $5.08" \times 5.08"$   \\ \hline
Polarization            & Circular              & Circular               \\ \hline
\makecell[l]{The distance between the \\ centers of the two guide stars}     &  15"  &  15" \\ \hline
\end{tabular}
\caption{The parameters of double guide star experiment of all solid-state laser}
\label{table: the parameters of double guide star experiment of all solid-state laser}
\end{table}

\begin{figure}[H]
\centering
\includegraphics[width=0.6\linewidth]{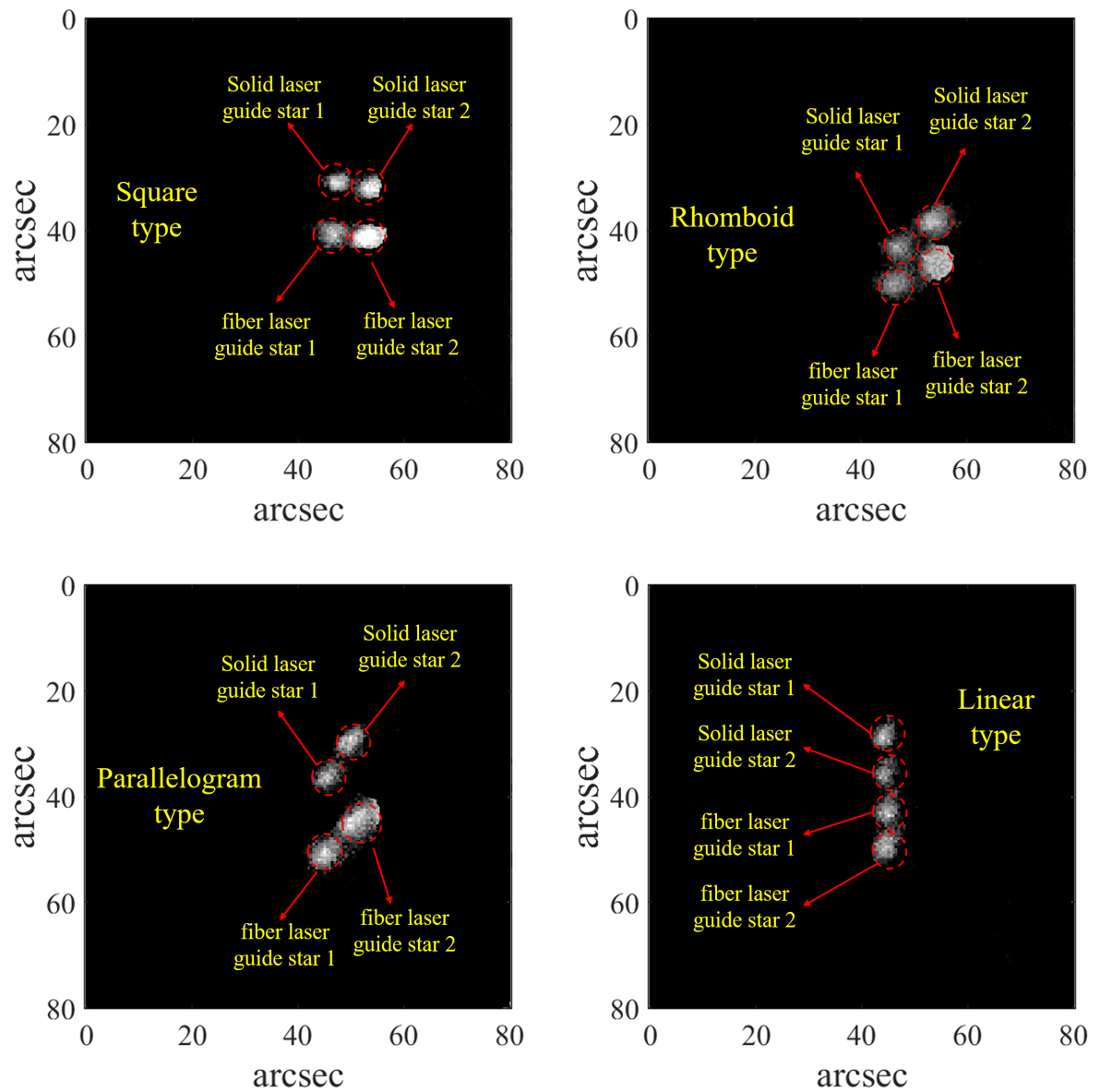}
\caption{The different type of laser guide star asterism}
\label{figure: the different type of laser guide star asterism}
\end{figure}

One single sodium laser can generate two guide stars, therefore, the laser guide star asterism including four guide stars could be formed by the all solid-state sodium laser and fiber sodium laser. And the shape of the laser guide star asterism can be changed by fine tuning the reflect mirror of the laser guide star facility. In the experiment, we changed the shape of the laser guide star asterism into Square, Rhomboid, Parallelogram and Linear. The results are shown in Figure \ref{figure: the different type of laser guide star asterism}. The output power of the all solid-state sodium laser and fiber sodium laser is 30W and the seeing is 2.5" when the experiment was conducting. The brightness and size of the guide stars and other parameters are summarized in Table \ref{table: the information of the laser guide star asterism}.Considering the instability of laser output power and the change of sodium column density (\citealt{Feng+etal+2021}), we have some uncertainties when measuring the brightness of different types of laser guide star asterism.

\begin{table}[H]
\scalebox{0.7}
\centering
\begin{tabular}{|l|l|l|l|l|}
\hline
\textbf{The type of LGS asterism}  &   \multicolumn{4}{|c|}{\textbf{Square type}}  \\ \hline              
\textbf{Parameter name}  & \textbf{Solid LGS 1}   &\textbf{Solid LGS 2}     &
\textbf{Fiber LGS 1}    &\textbf{Fiber LGS 2}                                      \\ \hline
Projected direction  &  Zenith               &  Zenith         &  Zenith          &  Zenith      \\ \hline
Brightness           & 8.46 Vmag             & 8.19 Vmag       & 8.72 Vmag        & 8.36 Vmag     \\ \hline
The Spot size on sky & $3.7" \times 3.9"$  & $3.9" \times 4.3"$ & $4.8" \times 4.5"$ & $3.8" \times 4.2"$ \\ \hline
\textbf{The type of LGS asterism}  &   \multicolumn{4}{|c|}{\textbf{Rhomboid type}}  \\ \hline               \textbf{Parameter name}                                                                         & \textbf{Solid LGS 1}                                                               &
\textbf{Solid LGS 2}                                                               &
\textbf{Fiber LGS 1}                                                               &
\textbf{Fiber LGS 2}                                                           \\ \hline
Projected direction  &  Zenith               &  Zenith         &  Zenith          &  Zenith      \\ \hline
Brightness           & 8.62 Vmag             & 8.27 Vmag       & 9.09 Vmag        & 8.76 Vmag     \\ \hline
The Spot size on sky & $3.8" \times 3.9"$  & $4.0" \times 4.2"$ & $4.1" \times 4.3"$ & $4.5" \times 4.3"$ \\ \hline
\textbf{The type of LGS asterism}  &   \multicolumn{4}{|c|}{\textbf{Parallelogram type}}  \\ \hline               \textbf{Parameter name}                                                                         & \textbf{Solid LGS 1}                                                               &
\textbf{Solid LGS 2}                                                               &
\textbf{Fiber LGS 1}                                                               &
\textbf{Fiber LGS 2}                                                           \\ \hline
Projected direction  &  Zenith               &  Zenith         &  Zenith          &  Zenith      \\ \hline
Brightness           & 8.65 Vmag             & 8.67 Vmag       & 9.13 Vmag        & 9.03 Vmag     \\ \hline
The Spot size on sky & $3.9" \times 3.7"$  & $3.8" \times 3.8"$ & $4.2" \times 4.0"$ & $3.7" \times 4.3"$ \\ \hline
\textbf{The type of LGS asterism}  &   \multicolumn{4}{|c|}{\textbf{Linear type}}  \\ \hline               \textbf{Parameter name}                                                                         & \textbf{Solid LGS 1}                                                               &
\textbf{Solid LGS 2}                                                               &
\textbf{Fiber LGS 1}                                                               &
\textbf{Fiber LGS 2}                                                           \\ \hline
Projected direction  &  Zenith               &  Zenith         &  Zenith          &  Zenith      \\ \hline
Brightness           & 8.53 Vmag             & 8.61 Vmag       & 8.73 Vmag        & 8.80 Vmag     \\ \hline
The Spot size on sky & $3.9" \times 4.3"$  & $4.2" \times 4.5"$ & $4.1" \times 4.4"$ & $3.9" \times 4.2"$ \\ \hline
\end{tabular}
\caption{The information of the laser guide star asterism}
\label{table: the information of the laser guide star asterism}
\end{table}

\section{Conclusion}
\label{section: conclusion}
A new laser guide star launching platform is demonstrated in this paper. The platform has two main functions. One is that the platform can compare the performance of sodium laser guide stars generated by two lasers with different parameters. In the field test, we calibrated the performance of the all solid-state sodium laser and the fiber sodium laser. When the power of the both lasers was adjusted to a maximum of 30W, we got a guide star with brightness of V7.87 mag and size on sky of $4.1” \times 4.2"$ generated by the all solid-state laser and a guide star with brightness of V7.81 mag and size on sky of $4.0” \times 4.2”$ generated by the fiber sodium laser. The other is that it can combine beams to generate a laser guide star, or split beams to generate multiple laser guide star asterism including four guide stars used by two lasers. And in the field test, we realized the generation of asterism with four laser guide stars, which the brightness of each guide star is 8-9 magnitude in V-band and the spot size on sky is 3.5”-4.5” when the power of the two lasers is 30W. This is the first time to realize the experiment of multi-laser guide star asterism in China. At present, conducted field tests were still not exhaustive due to the weather of the observatory, and the platform also has more room for improvement. In the future, more tests will be carried out on the platform, such as the improvement of laser beam quality, the test of maximum radius of generated asterism, the experiment of laser stability, etc.

\section{Acknowledgement}
\label{section: ackowledgement}
The research is partly supported by the National Natural Science Foundation of China (Grant Number 12173051,12173041,11733005),Youth Innovation Promotion Association,Chinese Academy of Sciences(No.2020376).

\end{document}